# Comparing fitness and drift explanations of Neanderthal replacement

Daniel R. Shultz[a,b]  Marcel Montrey[c]  Thomas R. Shultz[c,d]

McGill University

## Abstract

There is a general consensus among archaeologists that replacement of Neanderthals by anatomically modern humans in Europe occurred around 40K to 35K YBP. However, the causal mechanism for this replacement continues to be debated. Searching for specific fitness advantages in the archaeological record has proven difficult, as these may be obscured, absent, or subject to interpretation. Proposed models have therefore featured either fitness advantages in favor of anatomically modern humans, or invoked neutral drift under various preconditions. To bridge this gap, we rigorously compare the system-level properties of fitness- and drift-based explanations of Neanderthal replacement. Our stochastic simulations and analytical predictions show that, although both fitness and drift can produce fixation, they present important differences in 1) required initial conditions, 2) reliability, 3) time to replacement, and 4) path to replacement (population histories). These results present useful opportunities for comparison with archaeological and genetic data. We find far greater agreement between the available empirical evidence and the system-level properties of replacement by differential fitness, rather than by neutral drift.

## Introduction

Explaining the disappearance of Neanderthals from the archaeological record during the Upper Palaeolithic is a longstanding and ongoing debate involving anthropologists, archaeologists, biologists, and geneticists. *Homo neanderthalensis,* commonly known as Neanderthals, were an archaic branch of the genus Homo. They appeared first in Europe around 400K YBP, evolving out of ancestral variants of *Homo erectus* or *Homo heidelbergensis*, who had previously spread into Europe from Africa by 800K YBP. After being the sole hominin occupants of Europe for some 350K years, Neanderthals disappeared from the archaeological record at roughly the same time that anatomically modern humans (hereafter *Moderns*) spread from Africa through Europe via the Levant around 50K to 35K YBP [1–3].

Evidence for the replacement of Neanderthals by Moderns is morphological, archaeological, and genetic. Neanderthal skeletal remains feature unique morphological characteristics beyond the variation present in Moderns [4,5], and these traits disappear from the fossil record over time. Likewise, associated archaeological cultures, such as the Mousterian, are replaced by those associated with Moderns [6]. Genetic research in the last decade has revealed a Neanderthal DNA contribution of 1 to 4% to contemporary non-African populations, leading most experts to agree that some interbreeding did occur [7–10], although there is an alternative view that shared DNA was the result of common ancestors and ancient population divisions in Africa [11–13]. Interbreeding would indicate that Neanderthal genetics were eventually swamped by newly arrived Moderns. By extension, overlap in Modern expansion and Neanderthal disappearance was sufficient to permit temporary co-existence and contact.

---

[a] Department of Anthropology
[b] Department of History
[c] Department of Psychology
[d] School of Computer Science

Disputed elements center on issues of chronology and the replacement mechanism. Estimates of contact between Neanderthals and Moderns until the last Neanderthal occupation range from 50K to 30K YBP. Dating controversies cloud the last identified Neanderthal occupations in the Iberian Peninsula. Some researchers date these sites to around 30K YBP [14–16], while more recent research argues this is due to contamination with modern carbon, and suggests revised dates around 41-42K YBP for the youngest Neanderthal sites [2,17].

Proposed mechanisms for Neanderthal replacement can be divided into fitness- and neutral drift-based explanations. Fitness arguments have proposed a number of different advantages Moderns may have held over Neanderthals, including cognitive ability [18–21], hunting and technological prowess [22–28], diversity of diet and resource extraction [24,29], social cooperation [30], capacity for robust speech [31,32], and favorable birth/death rates [33].

More recently, simulation models invoking drift have been proposed, purporting not to require a fitness difference to achieve replacement. These explanations include fitness-neutral competitive exclusion via increased migration rate and/or greater Modern population size [34], and swamping or absorption of Neanderthal DNA via interbreeding with more numerous Moderns [35]. Drift models are supported by a line of archaeological research that rejects cognitive, technological, and cultural distinction of Neanderthals and Moderns [5,36–38]. Importantly, drift models require particular initial conditions to function (such as differential migration rates and starting population size differences), which could be undeclared proxies for a Modern fitness advantage. Similarly, fitness explanations remain a matter of debate because the archaeological record is equivocal regarding signatures of the foregoing proposed Modern fitness advantages [37]. Theoretically, small fitness advantages may be invisible archaeologically, particularly in similar complex organisms like Neanderthals and Moderns. A small net fitness difference in these cases could be the combined effect of hundreds of minute genetic features, difficult to identify with limited knowledge and data. For further discussion of how to define and study fitness and drift in the context of Neanderthal replacement, see supplemental information section 1.

We explore an alternative approach to this problem. We model replacement events abstractly, to determine whether fitness and drift replacement have different system-level properties, which could potentially be reflected in the empirical record of replacement. It would then be possible to use these signatures to identify fitness or drift as the more likely driver of a particular replacement event while remaining agnostic regarding what specific fitness advantages were (or were not) operative. Stochastic simulations are a critical tool for this approach, as 1) we are certain whether fitness or drift is operative, because we control the relevant parameters, and 2) we can experiment over many iterations and full ranges of parameter settings, generating an extensive dataset that allows us to determine the significance of each parameter to the outcome.

We examine four potentially different properties of fitness- and drift-based replacement. First is reliability: Over multiple iterations, at a given combination of parameter settings, how often does replacement in a particular direction occur? Second, related to reliability, are necessary initial conditions: What parameter settings are necessary to achieve a sufficiently reliable probability of replacement in a particular direction? Third is time to replacement: How long does replacement take under different parameter settings using fitness or drift? Fourth is the path to replacement: Do population histories differ significantly when fixation is driven by fitness or drift? In the case of Neanderthal replacement, each of these potential differences could leave archaeological and genetic signatures that are much more visible than the presence or absence of specific fitness differences. More broadly, analyzing systemic properties rather than specific traits potentially enables us to distinguish the action of fitness and drift in a wide variety of species replacement events.

## Model

Our Stochastic Bi-directional Stepping-stone (SBS) model integrates the study of fitness, drift, and both relative and total population sizes in explaining Neanderthal replacement. These factors are assessed by measuring the probability of a species reaching fixation and the number of time cycles needed to reach fixation. The agents in our model represent hominin bands of unspecified size. We use a stepping-stone configuration [39], similar to other recent approaches [34], with Neanderthal bands initially placed to the left and Modern bands to the right in the chain of stepping stones. A schematic representation of the model is presented in Figure S1.

Updating in our model follows a *death-birth* scheme [40]. In every time cycle, each band has a .01 chance of dying. Whenever a band dies, it is replaced by a replicate of an adjacent band. In most cases, a dying band is between two bands of the same type, in which case replacement does not move the inter-species border. If the neighbours of a dying band are of two different types, then the border position has a chance of moving one step to the left or right. The position of the border can change only if the replacement is from a different species, in which case there is a competition for the vacated position between the two immediate neighbours.

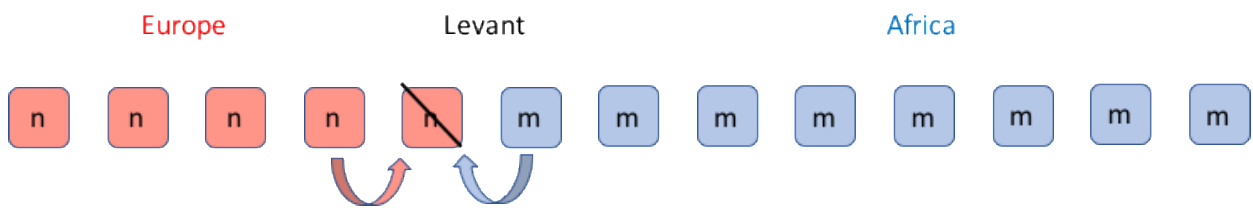

Figure S1. Schematic depiction of our Stochastic, Bidirectional, Stepping-stone (SBS) model. Neanderthal bands are labeled *n*, Modern bands *m*. The European continent is indicated with a red background, Africa with a blue background. The Levant is represented by the initial border, a dying band by a black cross-out, and competition for that position by curved arrows in the color of the replicate bands.

In these competitions, the replacement is *m* with probability equal to the Modern fitness parameter, and *n* otherwise (i.e., Neanderthal fitness is always 1 minus Modern fitness.) In a condition where Modern fitness is .5, border change is entirely due to neutral drift. If Modern fitness is greater than .5, the replacement is more likely to be *m*. Conversely, if Modern fitness is less than .5, the replacement is more likely to be *n*. Because total population size remains constant, even if border location changes, the model exhibits a *Moran-process*, a well-known device for simulating both neutral drift and natural selection by differential fitness [41].

## Results

Our stochastic simulation results are organized in terms of relative fitness, relative population size, total population size, border tracking and incursions, and band lifespan. Stochastic simulations are replicated 1000 times per parameter setting. Each simulation is run until fixation is reached. All simulation results are then plotted against analytical predictions derived from random walks (see supplemental information section 4). Throughout our figures, circles indicate stochastic simulation means, while solid curves represent analytical predictions.

**Effects of differential fitness**

To examine the main effect of differential fitness, we run our SBS model across a range of fitness values from .45 to .55, starting with 50 Neanderthal and 50 Modern bands. At neutral drift, the average Modern fixation probability is basically a coin flip (Figure 2). With fitness differences away from this neutral zone as small as .01, direction of fixation is strongly determined by whichever population is fitter. Fitness differentials of .02 away from pure drift make it virtually certain that the fitter population fixates. Analytical predictions show that, when initial population proportions are equal, this relationship has a sigmoidal shape, with the steepness of the curve controlled by the total number of bands (Equation S3). As the number of bands increases, so too does the probability of the favored type reaching fixation, and thus the steepness of the curve. Meanwhile, cycles to reach fixation are much greater with neutral drift and decrease rapidly as fitness deviates from .5.

This first simulation establishes that our model is truly bidirectional because it includes conditions under which Neanderthals clearly fixate (when Modern fitness is below .5). It also presages a recurring theme throughout our subsequent results: that a fitness advantage ensures relatively quick fixation, while neutral drift is both uncertain about fixation direction and slow to achieve fixation.

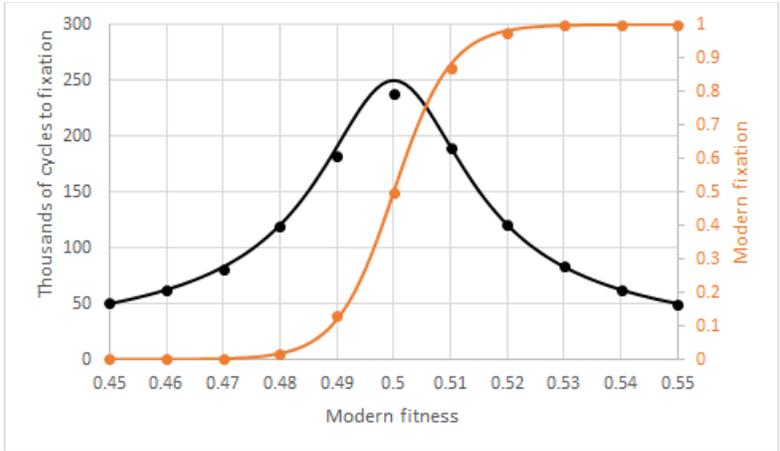

Figure 2. Fixation direction (in orange) and cycles to fixation (in black) as a function of Modern fitness. With a bit of distance from neutral drift (Modern fitness of .5), the fitter population is very likely to fixate.

**Effects of differential population sizes**

To study the effects of differential initial population sizes, we run our SBS model with population ratios ranging from 3:7 to 7:3 (100 bands total), across three levels of Modern fitness (.45, .5, and .55). Figure 3a shows that direction of fixation is affected by plausible differences in initial population size only with neutral drift (Modern fitness of .5). In fact, analytical findings show that the probability of a type reaching fixation is precisely equal to the initial proportion of bands of that type (Equation S4). With fitness deviations from pure drift as small as .05, differential fitness is decisive for direction of fixation and unaffected by initial population sizes within this parameter range. As in the previous experiment, the fitter population almost always fixates.

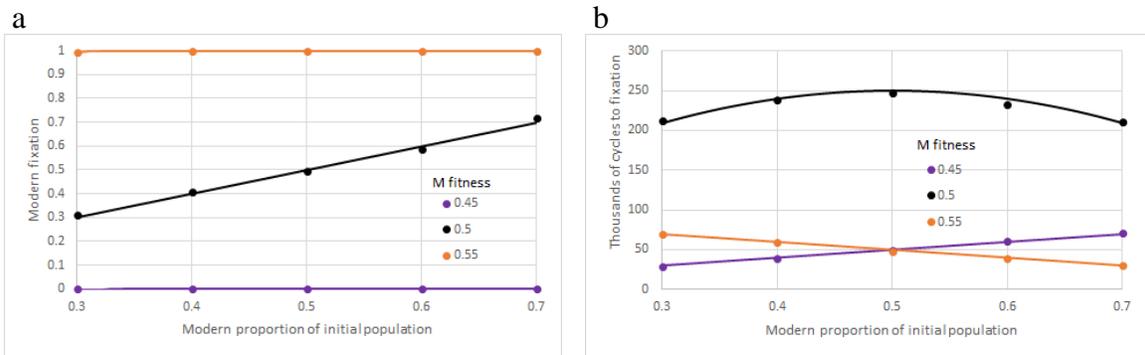

Figure 3. Effects of Modern proportion of initial population and Modern fitness on: a) Modern fixation, and b) time cycles to fixation.

Figure 3b shows that fixation takes much longer with neutral drift (Modern fitness of .5) than when there is a difference in fitness. This is because, at pure drift, the expected number of cycles to fixation scales with the product of the number of bands of each type (Equation S11). With relatively small fitness differences of .45 versus .55, fixation is much faster than drift, even if the extinguished population is initially more frequent. It is also apparent, from the slopes of the orange and purple lines, that fixation is even faster when the favored population has an initial advantage.

**Effects of total population size**

To assess the effects of total population size, we run our SBS model on four different numbers of total bands (100, 200, 300, and 400), across three levels of Modern fitness (.45, .5, and .55). Direction of fixation is plotted in Figure 4a as a function of fitness and number of bands. Although fixation direction is unaffected by number of bands, it is again strongly determined by differential fitness. As expected, with neutral drift, direction of fixation hovers around .5.

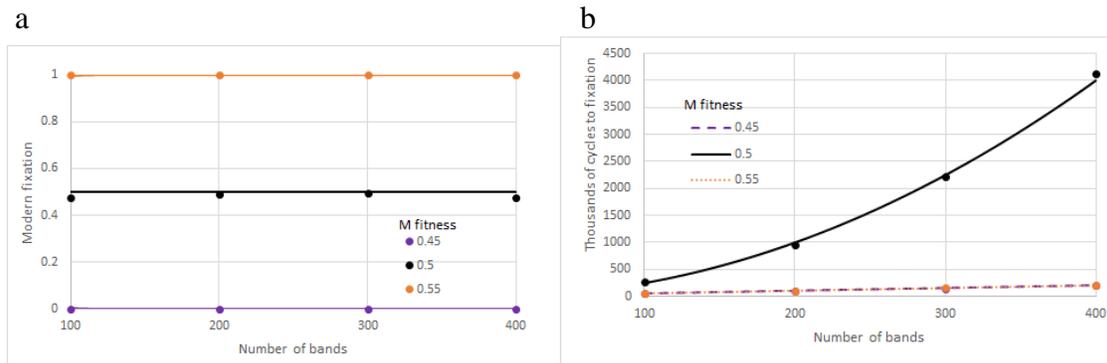

Figure 4.  Effects of number of bands and Modern fitness on: a) direction of fixation, and b) time cycles to fixation.

Time to fixation is plotted in Figure 4b, which shows a strong interaction between fitness and number of bands. Under neutral drift, when initial populations are equal, cycles to fixation increases as the square of half the total population (Equation S11). With a fitness differential, cycles to fixation grow as a linear function of total population (Equation S7-S10). At every level of population, neutral drift is again much slower to achieve fixation than differential fitness is.

**Border tracking and incursions**

We record population histories by tracking movement of the border over time cycles, from the initial border (defined by differential population sizes) to eventual fixation. Section 2 of supplemental information shows examples of such border movements from individual runs and how they can be used to assess the depth and duration of incursions of Neanderthal bands across the initial border. To efficiently quantify incursions, we use an incursion index, computed as the sum of incursion distances across the initial border at each time cycle. This measure integrates spatial distance past the initial border over time spent there.

To systematically explore the effect of fitness on Neanderthal incursions, we examine Modern fitness values from .5 up to .9. Initial populations are 50 Modern bands and 50 Neanderthal bands. Mean incursion scores drop sharply with small increases in Modern fitness (Figure 5a), indicating that Neanderthals go from making deep, repeated incursions into Modern territory at drift, to making almost no incursions past the initial zone of contact when Moderns have a fitness advantage. This can also be seen in Figure 5b, which shows the amount of time spent by Neanderthals in Modern territory, given that Moderns reach fixation (i.e., excluding replications where Neanderthals reach fixation). At drift, Neanderthals spend a great deal of time in Modern territory, despite their eventual replacement. Conversely, when Moderns have a fitness advantage (here a fitness of .55), Neanderthals spend almost no time in territory initially occupied by Moderns.

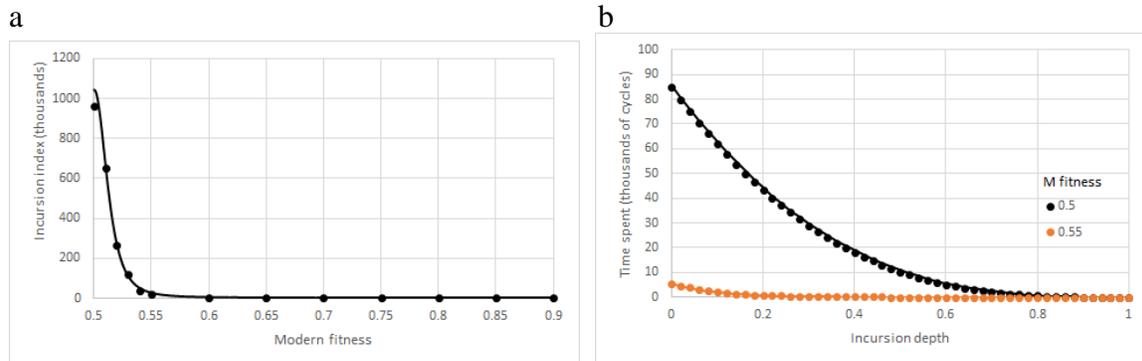

Figure 5. Effects of Modern fitness on: a) number, depth, and duration of incursions, and b) the amount of time spent by Neanderthals in Modern territory despite Moderns reaching fixatation.

**Band lifespans**

Because average band lifespan varies across parameter settings, we run a simulation experiment to determine whether these lifespans are plausible. Although it can be problematic to match simulation time with empirical time, this can be approximated by matching the start and end points of the process of interest for empirical and simulated times and then dividing simulation cycles by the approximate empirical time duration of the process being studied [42]. Most of the empirical data reviewed earlier suggests that Neanderthal replacement required about 10K years of contact with Moderns, although some estimate a much lower time frame of 2.6K to 5.4K years [2]. To be conservative, we divide cycles to fixation by 10,000 to estimate cycles per year. Then we calculate band lifespan in years by dividing band lifespan in cycles by cycles-to-fixation. Our band lifespan in cycles is 100 at a deathrate of .01. For example, using data from Figure 2, with Modern fitness of .55, the simulation requires about 50K cycles to fixation. Cycles per year is then about 50,000/10,000 = 5, and band lifespan in years is thus about 100 / 5 = 20 years. This contrasts with 250K cycles to fixation at pure drift, yielding about 250,000 / 10,000 = 25 cycles per year, and a band lifespan of about 100 / 25 = 4 years, not even enough time for a hominin to reach maturity. By these calculations, replacement with a band lifespan of 20 years at pure drift would require a continuous period of interaction and competition between Neanderthals and Moderns lasting at least 50,000 years.

To study band lifespan more systematically, we examine three levels of Modern fitness (.4, .5, and .6) where initial Modern population ratios range from 1:9 to 9:1. Mean band lifespans are plotted in Figure 6. The dotted line represents a band lifespan of 20 years, about one generation. Even a small fitness differential ensures that bands would endure long enough to produce offspring. However, at pure drift, band lifespans are too short for reproduction. Also, band lifespan scales with the inverse of the number of cycles. So when the population advantage for one species increases, the average number of cycles decreases, and lifespan increases. If fitness is operating in the same direction, then the whole process speeds up considerably and average band lifespan gets very high.

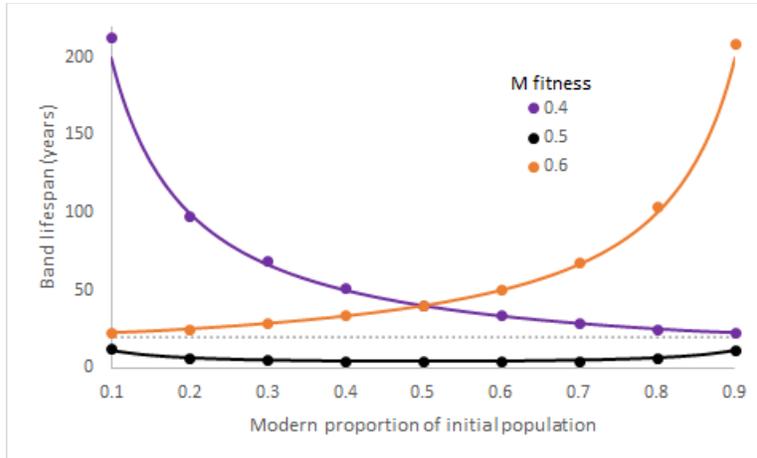

Figure 6. Mean band lifespan in years at three levels of Modern fitness and across initial Modern population ratios ranging from 1:9 to 9:1. The dotted line represents a band lifespan of 20 years.

## Discussion

Our results indicate four broad property differences between fitness and drift explanations: reliability, initial conditions, time to replacement, and path to replacement. We discuss how each of those differences favors a fitness explanation for Neanderthal replacement.

**Reliability**

With equal starting population size and no fitness difference, replacement of Neanderthals by Moderns occurs in about half of our simulations (Figure 2). On our planet, this replacement occurred only once. This still allows a decent probability of Neanderthal replacement (0.5) by purely neutral means. However, Moderns came into contact with archaic Africans, Neanderthals, and Denisovans. In each case, replacement by Moderns was the end result. Probability of consistent direction of replacement over three cases by chance alone is low ($.5^3 = 0.125$). This does not rule out pure drift as a replacement mechanism, but makes it a less likely explanation than fitness, where even a small advantage can lead to near-certain fixation.

**Initial conditions**

The reliability of a drift explanation increases above .5 with certain pre-conditions in place, namely a larger starting Modern population (Figure 3a). Previous research has shown that other conditions like unidirectional migration (Moderns can diffuse into Neanderthal territory but not vice versa) and migration rate (a greater probability Moderns will diffuse into Neanderthal territory than vice versa) drastically increase likelihood of Modern fixation [34]. However, we believe that unilateral migration and differential migration rate are undeclared proxies for a fitness advantage. It is otherwise difficult to explain why one species could diffuse successfully into the territory of another, but not the reverse.

Achieving Neanderthal replacement via drift with a reliability of .7 requires a 7:3 initial population ratio in favour of Moderns (Figure 3a). We recognize the theoretical possibility of complex environment-based arguments explaining why the Modern population could be so much larger despite equal fitness. Nevertheless, it is presently difficult not to view such a ratio as another fitness proxy, given the intimate relationship between biological fitness and replication frequency.

**Time to replacement**

Pure neutral drift takes far longer than fitness to achieve replacement (Figures 2, 3b, 4b). Pointedly, pure drift takes so long to achieve fixation that it requires such a high rate of band turnover as to be implausible for sustaining hominin bands long enough to produce a single generation of offspring (Figure 6). This is strong evidence that drift is too slow a mechanism to have caused Neanderthal replacement within the empirically observed time frame.

Theoretical work examining selection in the context of drift often recognizes a tradeoff between strength of selection and the speed at which fixation or evolutionary stability is reached [43,44]. Neutral drift is always operative and shifts frequencies of traits, organisms, and species, but may rarely result in total fixation at the species level, given that the process likely takes longer than stable ecological systems and equal fitness ratios tend to last. This is perhaps also why the biological sciences literature on extinction due to competitive replacement focuses mainly on fitness as the causal variable [45–47].

**Path to replacement**

Our experimentation also indicates that the path to species replacement is quite different for fitness versus pure drift (Figure 5). This also explains the increased time to replacement under pure drift – population sizes of both species expand and contract many more times before random chance results in fixation of one or the other. Our results demonstrate that drift replacement often results in numerous, long-lasting incursions of the replaced species deep into the initial territory of the replacer before the former finally goes extinct. This is a major property difference with potential archaeological and genetic implications. Under pure drift, our SBS model predicts an archaeological and genetic signature of Neanderthal artefacts, fossils, and DNA reaching into Africa, dated after initial contact with Moderns in Europe or the Levant. Current archaeological and genetic data do not seem to support this prediction. Modern sub-Saharan Africans do not feature Neanderthal DNA, and no Neanderthal artefacts or fossils have been found on the African continent. By contrast, recent research has identified genetic signatures of Modern migration from Europe to the Horn of Africa region, via the Levant, around 30K YBP [48]. This indicates there was no environmental barrier preventing hominins from moving in this direction, and suggests that a lack of Neanderthal diffusion into Africa was due to a fitness disadvantage in these new environments.

We acknowledge research attributing early fossils in the Levant to Moderns, with dates ranging from 90 to 180K YBP, predating Neanderthal occupation of that region [49–52]. Furthermore, a recent genetic study has proposed that interbreeding between Neanderthals and hominins more closely related to modern humans occurred in Europe as early as 220K YBP [53]. It is therefore possible that some early Moderns or proto-Moderns ventured out of Africa into the Levant, before dying off or being absorbed into the local Neanderthal population. It is also possible that any fitness advantages in favour of Moderns were not developed in these earlier periods. Until further empirical research clarifies these findings, we maintain our focus on the period of clearly verified Modern diffusion across Europe and interaction with Neanderthals between roughly 50 and 35K YBP. Genetic research using mtDNA to form Modern population estimates also supports the traditional chronology of around 40K YBP as being the first large scale wave of Modern diffusion into Europe [54]. For this period, a spatial, chronological, and genetic reading of the current empirical evidence suggests a path to Neanderthal replacement that our simulation and mathematical results argue is consistent with fitness explanations. Pointedly, the pure drift hypothesis predicts exactly the wrong result: evidence of Neanderthal presence in Africa.

It is important to note that Modern fitness advantage does not imply cognitive, physical, or moral superiority, but rather increased replication frequency. Replacement of Neanderthals by Moderns does not rank Neanderthals below Moderns on a hierarchical order of the genus Homo based on *how evolved* or cognitively complex they are. This would be a misapprehension of Darwinian evolution based on earlier strains of nineteenth century evolutionary theories characterized by progressive linearity and racialism. It would be similar to the erroneous but still-widespread notion that humans evolved from chimpanzees or other present-day apes, rather than seeing a variety of well-adapted contemporaneous species sharing common ancestors. For a broader discussion of our position on speciation, interbreeding, competitive exclusion, and genetic assimilation applied to Neanderthals and Moderns, see supplemental information section 3. Theoretically, there is no barrier to Neanderthals having equal or more sophisticated cognition, technology, culture, art, and symbolism, yet still being replaced by Moderns due to a slight fitness disadvantage invisible to us, but "visible to natural selection" [5]. Conversely, there is no particular reason to rule out cognitive differences as a potential mechanism.

On this note, a number of studies have identified Neanderthal genetic signatures suggesting raised extinction risk, including low genetic diversity, small population numbers, longer reproductive periods, and a higher coefficient of consanguinity, relative to Moderns [9,55–63]. Interestingly, lower Neanderthal fitness was quantified in a recent genetic simulation reporting that Neanderthals were .63 as fit as Moderns using additive (i.e., dominant) mutations [64]. With recessive mutations, Neanderthals were only .39 as fit as Moderns. A bit of algebra with these two values converts them to a Modern fitness advantage of .61 and .72, respectively, on our scale. At these levels of Modern fitness advantage, we find that Modern fixation is extremely rapid, virtually certain, leads to very few Neanderthal incursions past the initial zone of contact, and allows for realistically long band lifespans (see Figures 2-3, 5, 6, S5-S7, and S9).

## Conclusion

Although our reading of the current empirical record strongly suggests that a fitness explanation best covers current evidence, archaeological, genetic, and environmental research continually modify and update the empirical record. Future readings of this record may reach a different conclusion than we present here.

Our paper clearly acknowledges that Neanderthal replacement/genetic swamping, as seen in the empirical record, can be achieved by neutral drift, and we lay out the necessary conditions for this to occur. These are 1) a population size advantage for Moderns due to exogenous factors unrelated to fitness; 2) a small enough Neanderthal population for the population size effect to be in accordance with the empirically observed replacement length of about 10K years (or a continuous period of interspecies competition and interaction of at least 50K years); and 3) post-contact Neanderthal archeological sites in sufficient depth and density beyond the initial contact zone with Moderns. However, none of these preconditions are currently reflected in the empirical record.

Our simulation and analytical results instead consistently favor a differential fitness explanation for Neanderthal replacement, whether examining fixation direction, initial population differences, time to replacement, or path to replacement. Finally, our systematic experimentation with drift and fitness explanations under wide parameter ranges provides a useful framework for future attempts at explaining Neanderthal replacement and other replacement events.

## Acknowledgments

We are grateful to Michael Bisson and Graham Bell for helpful comments on a previous draft of this manuscript. This work was supported by a grant to TRS from the Natural Sciences and Engineering Research Council of Canada, RGPIN 7927-12.

## Author Contributions

DRS and TRS conceived the project. TRS and DRS executed early stochastic simulations. MM designed and wrote the analytical model and executed the final stochastic simulations. DRS, MM, and TRS wrote the manuscript.

# Supplementary Information

This file contains supplementary information on:
1. Defining fitness and drift
2. Individual simulation runs illustrating border tracking and incursion
3. Speciation, competitive exclusion, interbreeding, and genetic assimilation
4. Analytical predictions using random walks

## 1. Defining fitness and drift

Biological fitness is determined by the frequency at which the unit of study (alleles, traits, organisms, and average frequencies in populations and species) replicates itself in subsequent generations [1]. A unit producing more copies has higher fitness than a similar unit in the same environment producing less. Fitness is thus relative, changing with the environment and the existence of other units. For this reason, climate change models, sometimes proposed as a factor in Neanderthal extinction [2–4], should properly be classified as fitness models. Environmental shifts affecting the viability of Neanderthals, but not Moderns, ultimately indicate a Modern fitness advantage in the new environment. Upon identifying a unit replicating with increasing frequency relative to competitors, there are two possible explanations: 1) the unit has higher fitness, or 2) replication frequency is increasing due to drift, via random sampling. In the former, higher fitness means the unit undergoes natural and/or sexual selection more successfully than competitors. In the latter, fitness difference is absent, and the process is stochastic. In the real world, neutral evolution (drift) is always operative as part of the landscape in which natural selection occurs. Rigorous theoretical modeling experiments have demonstrated that under both fitness and drift (due to random sampling), when units are competing for finite replication locations, there is eventual fixation on one type [5,6]. This is relevant for closely related species like Neanderthals and Moderns, exploiting similar niches in an environment with a finite carrying capacity. Therein lies a problem of equifinality, as both fitness and drift can lead to an identical end result, fixation (i.e. replacement).

One solution to distinguish whether fitness or drift is responsible for a replacement event is to determine what fitness advantage the replacing unit may have had. If one can be identified, fitness is the main driver – if not, the process is neutral. This is largely the path the Neanderthal replacement debate has taken. One problem with this approach is that fitness has modern cultural connotations of mental and/or physical superiority. However, biological fitness can be much more subtle, referring not to intelligence or physical prowess per se, but only to an increased probability of replication, whatever the cause. Theoretically, small fitness advantages may be invisible archaeologically, particularly in similar complex organisms like Neanderthals and Moderns. A small net fitness difference in these cases could be the combined effect of hundreds of minute genetic features, difficult to identify with limited knowledge and data. This makes it all the more difficult to ascribe a replacement event to fitness. Likewise, ascribing a replacement event to neutral drift requires justifying any preconditions (such as a larger starting population and asymmetrical migration rate and direction) as the result of exogenous environmental factors rather than fitness difference. This is also difficult to prove empirically. Unequivocally attributing a replacement event to either fitness or drift can therefore be difficult.

## 2. Individual simulation runs illustrating border tracking and incursion

This section provides some additional background on measuring border tracking and incursion across the initial border in our simulations. Figure S2 presents an example plot of a single run which starts with 33 Neanderthal bands and 67 Modern bands, and a Modern fitness advantage of .67. In these simulations, one band is randomly select to die at each time cycle. There is considerable variation in these border plots; what they have in common is that they are not monotonic.

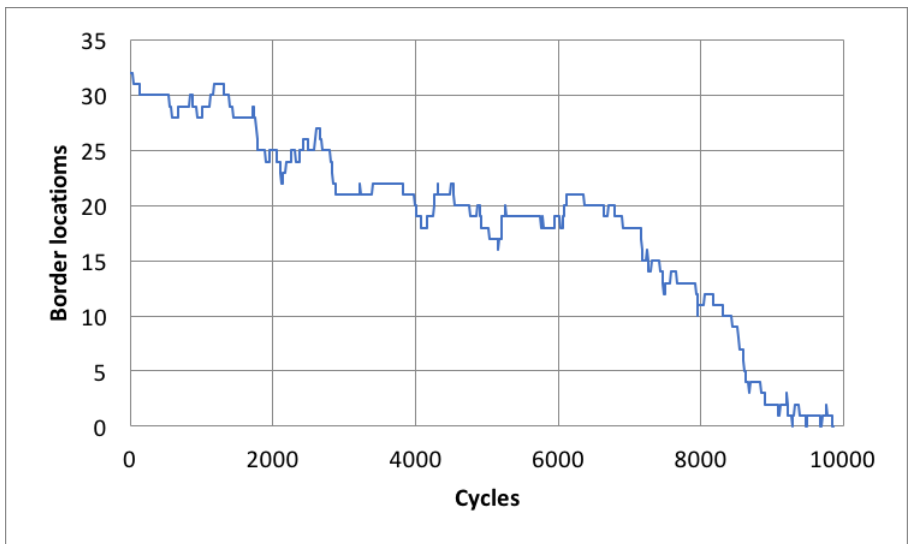

Figure S1. Example of border tracking in one replication with initial populations at 33 Neanderthal bands and 67 Modern bands, and Modern fitness of .67.

As noted in the main paper, such border tracking enables identification of incursions across the initial border between Neanderthals and Moderns. Our incursion index computes the sum of incursion distances across the initial border, at each time cycle. This integrates time frequencies and spatial distance spent across the border. Two examples of replicate incursions are presented in Figures S2 and S3, with the initial border represented by a horizontal red line. Both example simulations start with 25 Neanderthal bands and 75 Modern bands. The plot in Figure S2 is based on neutral drift, with a Modern fitness of .5; that in Figure S3 uses a Modern fitness of .6. Notice that there are considerably more and deeper Neanderthal incursions under neutral drift than with Modern fitness of .6.

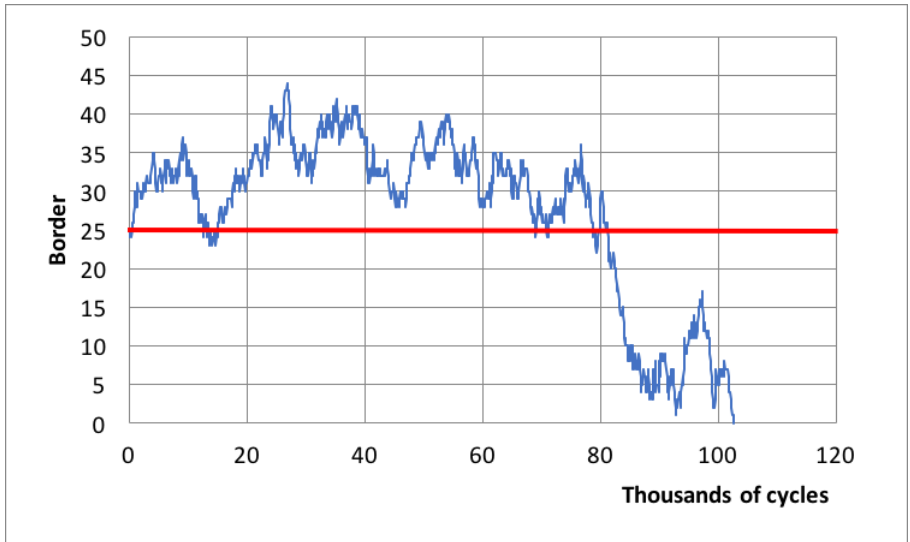

Figure S2. Example of border incursions in one replication with initially 75 Modern bands and 25 Neanderthal bands, with neutral drift (Modern fitness of .5). In this case, the Neanderthal incursion index registers 612,486 before eventual Modern fixation.

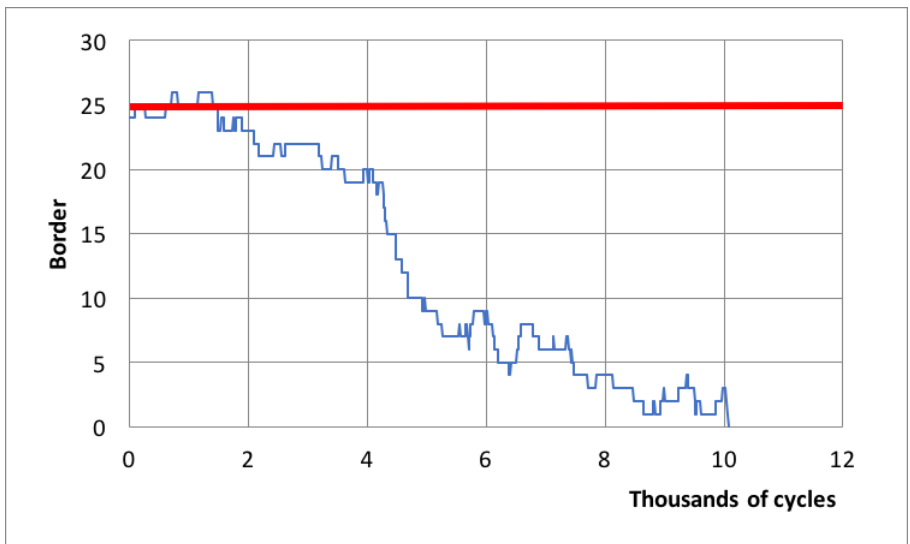

Figure S3. Example of border incursions in one replication with initially 75 Modern bands and 25 Neanderthal bands, with Modern fitness of .6. In this replication, the Neanderthal incursion index registers only 343 before Modern fixation.

Results in the main paper focus on a larger, more systematic experiment using these techniques to study paths to replacement.

### 3. Speciation, competitive exclusion, interbreeding, and genetic assimilation

Much of the Neanderthal replacement debate currently revolves around issues of speciation, regardless of whether this is made explicit. To some, the implication of interbreeding and Neanderthal genetic presence in modern Europeans is that the issue of replacement is misguided; Neanderthals and Moderns are populations of a single species.

Therefore, it is useful to examine the grounds on which Neanderthals are classified as a separate species.

Speciation is a critical process in biological evolution. Although defining species is somewhat arbitrary, the biological realities that classification seeks to understand are real [7]. Biological realities underlying species classification in sexually reproducing organisms are concerned with clusters of genotypic and phenotypic similarity, referred to as clades – groupings that include a common ancestor and all the descendants of that ancestor. These clusters arise out of accumulated mutations during a sufficiently long period of reproductive isolation, typically the result of geographic separation [1]. Members of the same species are more similar to each other within these clusters than to members of another species. Unlike populations within a species (such as modern humans), the average genetic and phenotypic difference between two species is generally greater than the standard deviation of distribution within the species. A great deal of research has identified morphological and genetic traits held in common amongst Neanderthals that lie outside the scope of modern human variation. For morphological traits, see [8–14]. For genetics, see [15–23].

It is important to note that the capacity for viable sexual reproduction is no longer used to define species in the biological sciences [1,7]. Wolves and coyotes, among many others, are well-studied cases of viable inter-species reproduction [24]. Applying contemporary, cladistic biological standards, it is reasonable to classify Neanderthals and Moderns as two different, albeit very closely related, species. Because species-level fitness can be calculated as the average replication frequency across species members, the probability that this average would be identical for two different species exploiting similar niches (Neanderthals and Moderns) seems very low. As our models demonstrate, only a slight deviation from equal fitness is sufficient to result in reliable and rapid species replacement.

Fitness-based competitive exclusion leading to replacement by an invasive and/or very closely related species is common and well-observed in the biological sciences [25–31]. Furthermore, competitive exclusion-based extinction between same-genus species frequently includes hybridization (interbreeding) and genetic pollution/swamping [31,32]. In other words, the presence of interbreeding does not, by itself, indicate pure drift. The assertion that Neanderthals are ancestors of modern humans, or are themselves modern humans, rather than an extinct branch of the same genus, is incorrect if made on the grounds of interbreeding alone. Neanderthals are ancestral and/or not extinct insofar as some of their genetic material persists, but this is inconsistent with consensus definitions of ancestral species and extinction in biology. A fitness difference leading to a combination of competitive exclusion and genetic pollution via hybridization would place Neanderthals replacement in the company of a multitude of well-observed species replacement events. Explicitly, an assimilation hypothesis for Neanderthal replacement [33,34], resulting in a small amount of Neanderthal DNA in the subsequent population, bears the signatures of replacement via differential fitness.

## 4. Analytical predictions using random walks

Our stochastic simulation model represents a series of discrete agents (i.e. bands) on a one-dimensional vector. However, because bands are represented solely as members of a given type (Moderns or Neanderthals), a meaningful change only occurs when a band is replaced by a band of another type. Although all band deaths and replacements are recorded, only movement of the border between Modern and Neanderthal populations is significant for species replacement. As a result, replacement can be represented as a random walk of this border. Throughout this paper, we plot simulation results against analytical predictions made on the basis of such random walks. To move between these analytical predictions and our SBS simulations, we need only convert time scales. We do so by observing that the border moves at each cycle with probability equal to death rate, $r = 0.01$.

In this section, we reinterpret group-level competition between Neanderthals ($N$) and Moderns ($M$) as a one-dimensional random walk of the border between the two types. Let $\omega_N$ be the fitness of Neanderthals and $\omega_M$ be the fitness of Moderns, where the fitness of each type falls between 0 and 1, and $\omega_N + \omega_M = 1$. Next, let $\lambda$ be the total number of bands and $b$ the position of the border between Neanderthals and Moderns, such that $0 \leq b \leq \lambda$. We take $b = 0$ as an absorbing state where Neanderthals have gone extinct and Moderns have reached fixation, and define $Pr(M)$ as the probability of reaching this state. This leaves $b = \lambda$ as an absorbing state where Moderns have gone extinct and Neanderthals have reached fixation, and we define $Pr(N)$ as the probability of reaching this state. Note that the position of the border $b$ also represents the number of Neanderthal bands in the population. The number of Modern bands is therefore $\lambda - b$.

This section is divided into four parts. In the first, we derive the probability of a given type reaching fixation, which is equivalent to the common derivation of the probability of reaching a given absorbing state in a random walk. In the second part, we derive the expected number of steps to fixation, which is equivalent to the common derivation of the expected number of steps to absorption in a random walk. In the third part, we derive the expected amount of incursion by Neanderthals into Modern territory following initial contact, which corresponds to the incursion index discussed in the main text. Finally, in the fourth part, we derive the expected amount of time Neanderthals spend in Modern territory, given that Moderns reach fixation.

**Probability of fixation**

Let $Pr(N|b)$ be the probability of Neanderthals reaching fixation, given that the border (or total number of Neanderthal bands) is $b$. At any $b$, $Pr(N|b)$ is given by the product of $Pr(N|b+1)$ and the probability of the border moving to the right ($\omega_N$), plus the product of $Pr(N|b-1)$ and the probability of the border moving to the left ($\omega_M = 1 - \omega_N$). This yields a recurrence relation

$$Pr(N|b) = (1 - \omega_N)Pr(N|b-1) + \omega_N Pr(N|b+1), \tag{S1}$$

which has a known family of solutions:

$$Pr(N|b) = c_1 \left(\frac{1}{\omega_N} - 1\right)^b + c_2. \tag{S2}$$

Neanderthals have gone extinct when $b = 0$, so $Pr(N|0) = 0$. Conversely, Neanderthals have reached fixation when $b = \lambda$, so $Pr(N|\lambda) = 1$. Substituting these values into Equation S2 creates a system of equations. Solving it gives the fixation probability:

$$Pr(N|b) = \frac{\left(\frac{1-\omega_N}{\omega_N}\right)^b - 1}{\left(\frac{1-\omega_N}{\omega_N}\right)^\lambda - 1}. \tag{S3}$$

We can make several observations. First, this function is sigmoidal over $0 \leq \omega_N \leq 1$. It has a single discontinuity at $\omega_N = 1/2$, where fitness of the two types is equal, and so we must derive a separate solution for the case of drift. Finally, as the total number of bands increases, so too does the function's steepness (growth rate between asymptotes).

The probability of fixation at drift is given by setting $w_N = 1/2$ in the initial recurrence relation (Equation S1), and then repeating the derivation:

$$Pr(N|b) = \frac{b}{\lambda}. \tag{S4}$$

Because the probability of Moderns reaching fixation is $1 - Pr(N|b)$, we observe that, at drift, the probability of a given type reaching fixation is equal to the initial proportion of bands of that type.

**Steps to fixation**

Let $s(b)$ be the expected number of steps (movements of the border) needed to reach fixation, when starting from position $b$. At any given $b$, $s(b)$ is given by the recurrence relation

$$s(b) = 1 + (1 - \omega_N)s(b-1) + \omega_N s(b+1), \tag{S5}$$

where 1 represents the single step needed to move from this position, $\omega_M = 1 - \omega_N$ is the probability of moving to the left, $s(b-1)$ is the expected number of steps to fixation from $b-1$, $\omega_N$ is the probability of moving to the right, and $s(b+1)$ is the expected number of steps to fixation from $b+1$. This recurrence has a known family of solutions:

$$s(b) = \frac{b}{1 - 2\omega_N} + \frac{\omega_N}{(1 - 2\omega_N)^2} + c_1 \left(\frac{1-\omega_N}{\omega_N}\right)^b + c_2. \tag{S6}$$

No more steps can be taken when a type reaches fixation, so $s(0) = 0$ and $s(\lambda) = 0$. Substituting these values into Equation S6 creates a system of equations. Solving it gives the expected number of steps:

$$s(b) = \frac{b}{1 - 2\omega_N} - \left(\frac{\lambda}{1 - 2\omega_N}\right) \left[\frac{\left(\frac{1 - \omega_N}{\omega_N}\right)^b - 1}{\left(\frac{1 - \omega_N}{\omega_N}\right)^\lambda - 1}\right]. \tag{S7}$$

How does this value grow with the total number of bands? There are two cases to consider. In the first, Modern fitness exceeds Neanderthal fitness, and so the fitness ratio $(1 - \omega_N)/\omega_N > 1$. This ratio gets very large when raised to the power of $b$ or $\lambda$, and so we can safely ignore the constant $-1$ in the bracketed term. That term then simplifies to

$$\left[\frac{\omega_N}{1 - \omega_N}\right]^{\lambda - b}, \tag{S8}$$

which rapidly approaches 0 as the number of Modern bands $(\lambda - b)$ gets large. As a result, only the first term in Equation S7 remains

$$s(b) \sim \frac{b}{1 - 2\omega_N}, \tag{S9}$$

and so the expected number of steps grows linearly with $b$, the initial number of Neanderthal bands. In the second case, Neanderthal fitness exceeds modern fitness, which means that the fitness ratio $(1 - \omega_N)/\omega_N < 1$. When raised to the power of $b$ or $\lambda$, this ratio rapidly approaches 0, causing the bracketed term to become $-1/-1 = 1$. Removing it from Equation S7 and then simplifying gives

$$s(b) \sim \frac{\lambda - b}{2\omega_N - 1}, \tag{S10}$$

and so the expected number of steps grows linearly with $\lambda - b$, the inital number of Modern bands. If we assume that the initial population ratio $(b/\lambda)$ remains constant, then in both cases the expected to number of steps grows linearly with the total number of bands.

To find the expected number of steps at drift, we set $w_N = 1/2$ in the initial recurrence relation (Equation S5), and then repeat the derivation:

$$s(b) = b(\lambda - b). \tag{S11}$$

We find that, at pure drift, the expected number of steps is a product of the number of bands of each type. Assuming an initially equal number of Neanderthal and Modern bands, the expected time to fixation is the square of one half the total number of bands. Note that, for both fitness and drift, converting from steps to simulation cycles merely requires dividing $s(b)$ by the death rate $r$.

**Incursion amount**

One empirically-relevant measure we propose to help distinguish between drift and fitness explanations is the amount of incursion into Modern territory that occurs once the two types meet. An incursion is characterized by both its depth into Modern territory and its duration. We calculate the total amount of incursion as a sum of each incursion's depth multiplied by its duration.

We begin by deriving $v(x|M)$, the expected number of times that a position $x$ in Modern territory will be visited by Neanderthals from a starting position $b$, given that Moderns reach fixation. If we assume that $x$ is visited exactly once before Modern fixation is achieved, then we can make a few observations. First, $b < x$, because Neanderthals occupy the left side of the vector while Moderns occupy the right. Second, $x$ must be reached before $0$, because $0$ is an absorbing state where Moderns have reached fixation. Third, after reaching $x$, the border must move left and never return to $x$, because moving right would imply crossing $x$ a second time to reach $0$. The probability of reaching $x$ from $b$ exactly once before Modern fixation is therefore

$$\frac{Pr(x|b)\omega_M \overline{Pr}(x|x-1)}{Pr(M|b)}, \tag{S12}$$

where $Pr(x|b)$ is the probability of reaching $x$ from $b$ before being absorbed by state $0$, $\omega_M$ is the probability of moving left from position $x$, $\overline{Pr}(x|x-1)$ is the probability of reaching absorbing state $0$ from $x-1$ without ever returning to $x$, and $Pr(M|b)$ is the probability of Modern fixation. The number of times we expect to visit $x$ once before Modern fixation is equal to the above probability:

$$v_1(x|M) = \frac{Pr(x|b)\omega_M \overline{Pr}(x|x-1)}{Pr(M|b)}. \tag{S13}$$

If the border reaches position $x$ twice before Modern fixation, then one of two additional things must happen before $x$ is left for good: Either the border must move left to $x-1$ (with probability $\omega_M$), and end up back at $x$ before Modern fixation; or the border must move right to $x+1$ (with probability $\omega_N$), and end up back at $x$ without Neanderthals reaching fixation. Therefore, once we are at $x$, the probability of reaching $x$ again is:

$$\omega_M Pr(x|x-1) + \omega_N Pr(x|x+1). \tag{S14}$$

Taken together, the probability of reaching $x$ from $b$ exactly twice before Modern fixation is

$$\frac{Pr(x|b)\omega_M \overline{Pr}(x|x-1)\bigl(\omega_M Pr(x|x-1) + \omega_N Pr(x|x+1)\bigr)}{Pr(M|b)}, \tag{S15}$$

which is just a product of the probability of reaching $x$ once before Modern fixation (Equation S12) and the probability of returning to $x$ once after having reached it (Equation S14). To get the expected number of visits to $x$ as a result of passing through $x$ exactly twice, we multiply the probability given in Equation S15 by the number of visits, two:

$$v_2(x|M) = \frac{2Pr(x|b)\omega_M\overline{Pr}(x|x-1)\big(\omega_M Pr(x|x-1) + \omega_N Pr(x|x+1)\big)}{Pr(M|b)}. \tag{S16}$$

From here, it is apparent that we can generalize this reasoning to any number of visits to $x$. The probability of reaching $x$ exactly $i$ times before Modern fixation is

$$\frac{Pr(x|b)\omega_M\overline{Pr}(x|x-1)\big(\omega_M Pr(x|x-1) + \omega_N Pr(x|x+1)\big)^{i-1}}{Pr(M|b)}, \tag{S17}$$

and the expected number of visits to $x$ as a result of passing through $x$ exactly $i$ times is:

$$v_i(x|M) = \frac{i Pr(x|b)\omega_M\overline{Pr}(x|x-1)\big(\omega_M Pr(x|x-1) + \omega_N Pr(x|x+1)\big)^{i-1}}{Pr(M|b)}. \tag{S18}$$

We can describe the total expected number of visits to state $x$ using an infinite sum over $i$, where we add up the expected number of visits to $x$ for every $i$:

$$\begin{aligned}v(x|M) &= \sum_{i=1}^{\infty} v_i(x|M) \\ &= \sum_{i=1}^{\infty} \frac{i Pr(x|b)\omega_M\overline{Pr}(x|x-1)\big(\omega_M Pr(x|x-1) + \omega_N Pr(x|x+1)\big)^{i-1}}{Pr(M|b)}.\end{aligned} \tag{S19}$$

This expression can be rewritten as

$$v(x|M) = \frac{Pr(x|b)\omega_M\overline{Pr}(x|x-1)}{Pr(M|b)} \sum_{i=0}^{\infty} (i+1)\big(\omega_M Pr(x|x-1) + \omega_N Pr(x|x+1)\big)^i, \tag{S20}$$

where the infinite sum has a nice solution:

$$\frac{1}{\big(1 - \omega_M Pr(x|x-1) - \omega_N Pr(x|x+1)\big)^2}. \tag{S21}$$

The expected number of visits to $x$ from $b$, before Moderns reach fixation, is therefore:

$$v(x|M) = \frac{Pr(x|b)\omega_M\overline{Pr}(x|x-1)}{Pr(M|b)\big(1 - \omega_M Pr(x|x-1) - \omega_N Pr(x|x+1)\big)^2}. \tag{S22}$$

To put Equation S22 in a usable form, we next solve for all of the conditional probabilities. For notational convenience, let $\gamma$ represent the fitness ratio $(1-\omega_N)/\omega_N$. First, we can infer from our solution for fixation probability (Equation S3) that the probability of reaching $x$ from $b$ is:
$$Pr(x|b) = \frac{\gamma^b - 1}{\gamma^x - 1}. \tag{S23}$$

Similarly, the probability of reaching $x$ from $x - 1$, without being absorbed by state $0$, is
$$Pr(x|x-1) = \frac{\gamma^{x-1} - 1}{\gamma^x - 1}, \tag{S24}$$

which entails that the probability of reaching state $0$ from $x - 1$, without ever revisiting $x$, is:
$$\overline{Pr}(x|x-1) = 1 - \frac{\gamma^{x-1} - 1}{\gamma^x - 1}. \tag{S25}$$

Finally, the probability of reaching $x$ from $x + 1$, without being absorbed by state $\lambda$, is:
$$Pr(x|x+1) = 1 - \frac{\gamma - 1}{\gamma^{\lambda - x} - 1}. \tag{S26}$$

Replacing these solutions (as well as the probability of Modern fixation) into Equation S22 and simplifying gives:
$$v(x|M) = \frac{(\gamma + 1)(\gamma^b - 1)(\gamma^\lambda - \gamma^x)^2}{\gamma^x(\gamma - 1)(\gamma^\lambda - 1)(\gamma^\lambda - \gamma^b)}. \tag{S27}$$

Finally, we can combine the expected number of visits to each state, $v(x|M)$, with the depth of every such incursion into Modern territory, $d(x)$. Note that the depth of an incursion is simply the distance between $x$ and the initial border $b$:
$$d(x) = x - b. \tag{S28}$$

Multiplying these values together gives the expected amount of incursion across the initial border $b$:
$$a(b) = \sum_{x=b+1}^{\lambda-1} v(x|M)d(x). \tag{S29}$$

Replacing in Equations S27 and S28, and then simplifying gives:
$$a(b) = \frac{(\gamma + 1)(\gamma^b - 1)}{(\gamma - 1)(\gamma^\lambda - 1)(\gamma^\lambda - \gamma^b)} \sum_{x=b+1}^{\lambda-1} \frac{(\gamma^\lambda - \gamma^x)^2(x - b)}{\gamma^x}. \tag{S30}$$

Solving the finite sum gives
$$\frac{\gamma^{2\lambda+1} + \gamma^{2b+1} - \gamma^{\lambda+b}((\gamma - 1)^2(\lambda - b)^2 + 2\gamma)}{\gamma^b(\gamma - 1)^2}, \tag{S31}$$

which we replace back into Equation S30, and then simplify to arrive at our final expression:

$$a(b) = \frac{(\gamma+1)(\gamma^b-1)\left(\gamma^{\lambda+b}((\gamma-1)^2(\lambda-b)^2+2\gamma)-\gamma^{2\lambda+1}-\gamma^{2b+1}\right)}{\gamma^b(\gamma-1)^3(\gamma^\lambda-1)(\gamma^b-\gamma^\lambda)}. \quad (S32)$$

However, when fitness of the two types is equal, this expression is undefined.

To get $a(b)$ in the case of drift, we substitute $\omega_N = \omega_M = 1/2$ into Equation S22, along with conditional probabilities inferred from Equation S4. This yields

$$v(x|M) = \frac{2b(\lambda-x)^2}{\lambda(\lambda-b)}, \quad (S33)$$

and:

$$a(b) = \frac{b(\lambda-b)(\lambda-b-1)(\lambda-b+1)}{6\lambda}. \quad (S34)$$

Note that, for both fitness and drift, converting from incursion amount to incursion index in the simulation requires dividing $a(b)$ by the death rate $r$.

**Time spent in Modern territory**

Let $t(x|M)$ be the expected number of steps Neanderthals will spend at position $x$ in Modern territory, given that Moderns reach fixation and that the initial border is at $b$. Note that when the border moves right to $x$, all territory left of $x$ is still occupied by Neanderthals. The amount of time spent by Neanderthals at $x$ is therefore not equal to the number of visits to $x$, $v(x|M)$, but rather to the total number of visits to any position right of $x$ (including $x$):

$$t(x|M) = \sum_{i=x}^{\lambda-1} v(i|M) \quad (S35)$$

Applying our solution for $v(x|M)$ from Equation S27, solving the finite sum, and then simplifying gives:

$$t(x|M) = \frac{(\gamma^b-1)((\gamma+1)(\gamma^{2\lambda+1}-\gamma^{2x})-\gamma^{\lambda+x}(\gamma^2-1)(2\lambda-2x+1))}{\gamma^x(\gamma^\lambda-1)(\gamma^\lambda-\gamma^b)(\gamma-1)^2} \quad (S36)$$

However, this expression is undefined when fitness of the two types is equal, and so we must repeat this process for drift by applying Equation S33 instead. Doing so, solving the finite sum, and then simplifying gives:

$$t(x|M) = \frac{b(\lambda-x)(\lambda-x+1)(2\lambda-2x+1)}{3\lambda(\lambda-b)}. \quad (S37)$$

Once again, converting from steps to simulation cycles requires dividing $t(x|M)$ by the death rate $r$.